# CHARGE TRAPPING IN FERROELECTRIC POLYMERS

A. F. Butenko

*Department of Physics and Materials Science, Odessa National Academy of Food Technologies, Odessa, Ukraine*

Experimental results are presented on anomalous behavior of absorption currents in PVDF during the stepwise increase of the voltage applied through a corona. These results, supplemented with the dynamics of the time constant of the electret potential decay are consistent with a hypothesis assuming a deep trapping of the charge carriers during the polarization buildup in ferroelectric polymers. The model calculations are made on a distribution of the potential energy at the surface of the polarized crystallites, proving that additional sites for the charge trapping are created there. It is shown how to provide the most favorable conditions for obtaining a stable polarization in ferroelectric polymers.

## 1. Introduction

There are two opinions on the role of the space charge in ferroelectric polymers. Majority agrees that existence of the space charge is essential [1-8]. However, others consider its role as negligible [9] and even believe that the space charge prevents obtaining high values of polarization [10]. In this respect, the following questions seem to be rether important. What are the sources of the space charge? What is origin of the traps for charge carriers and how do they appear? How localization and redistribution of the charge can be recognized and identified experimentally?

Answering the question one must rely on certain experimental facts and some fundamental principles. It is known that the polarized state of a ferroelectric monocrystal is unstable, because the depolarization field tends to return the crystal to its initial state with zero average polarization [11]. This field must be compensated (neutralized) anyhow in order to fix the new direction of the polarization. This is usually performed by real charges at the surface of the ferroelectric, if the sample is not short-circuited, or by the charges at electrodes. It is obvious that the processes in ferroelectric polymers are more complex than those in monocrystals, because crystallites of the former are dispersed in the paraelectric amorphous phase. The surface charge therefore can only neutralize the average field, but the local depolarization field in every crystallite can only be compensated by the charges localized at the crystallites boundaries.

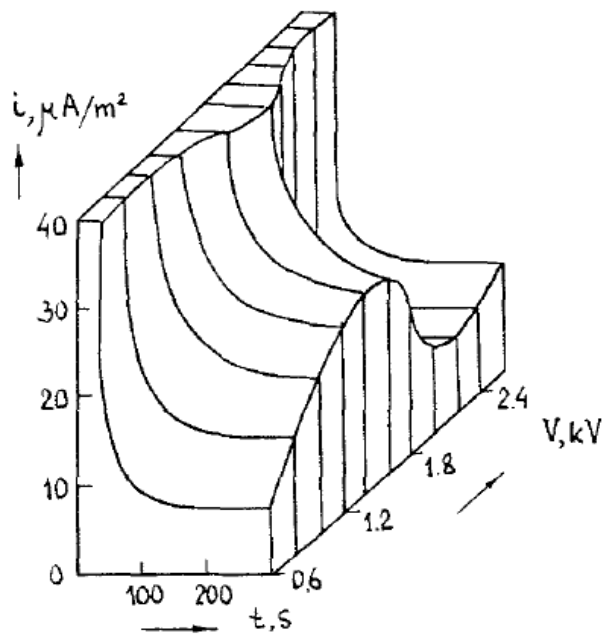

Fig. 1 Absorption currents in PVDF at constant voltages applied in steps of 200 V in a positive corona discharge

The hypothesis of the charge trapping related to polarization in PVDF was proposed by Pfister et al [2] who observed decrease in conductivity of the samples poled by the thermoelectret method. Eisenmenger et al [3-5] adopted this idea to explain polarization non-uniformity in PVDF and suggested that the charge trapping took place at the boundaries of the polarized zones. Fedosov [6] developed a model of the polarization buildup in a corona, in which an important role was attributed to the injected and trapped charge carriers. De Reggi and Broadhurst [7] have found that polarization in PVDF abruptly dropped to zero near electrodes, indicating that the injected space charge affected the spatial distribution of the field. An important role of space charge and depolarization phenomena was revealed by Ikeda et al [8]l in processes of switching in ferroelectric polymers.

In this paper, some additional results are presented on charge trapping related to the polarization buildup. They include the analysis of the anomalous absorption currents in PVDF and a quantitative estimation of the potential energy profile at a surface of the polarized crystallite.

## 2. Experimental procedure

The study was performed on 25 μm thick uniaxially stretched PVDF films containing amorphous and crystalline (Form 1) phases. The samples were placed in a corona triode with the charging voltage controlled by the grid potential. The voltage was applied in steps of 200 V in a range of 600-2800 V. The absorption current was recorded for 15 min at every applied voltage. Then the voltage was switched off for 5 min and the decay of the electret potential was monitored before the next step was commenced.

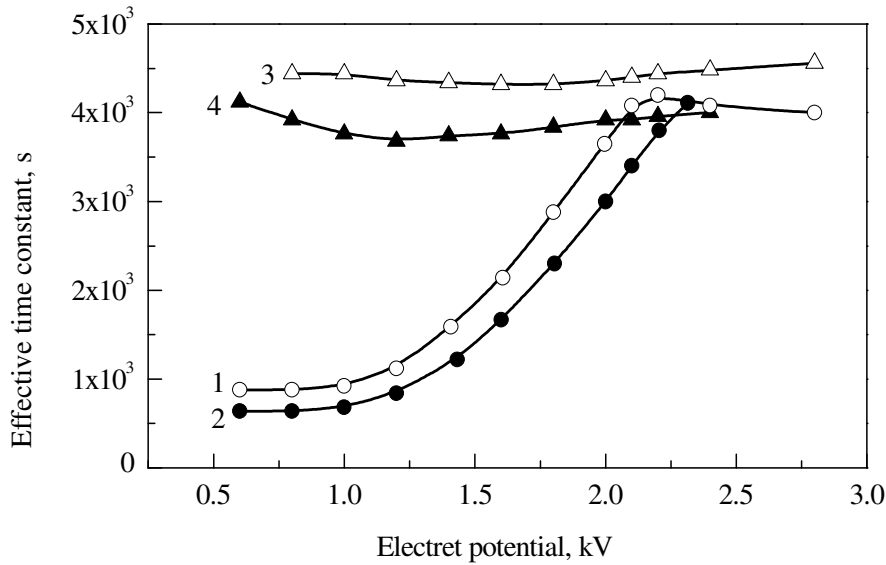

Fig. 2 Effective time constant of the electret potential decay during first (1,2) and second (3,4) poling in positive (1,3) and negative (2,4) corona discharge.

Absorption currents at different voltages are shown in Fig. 1. The anomaly is in a non-monotonicity of the current growth with increasing of the applied voltage. Considerable decrease of the current is observed at 1.6-2.0 kV indicating that the apparent conductivity also decreases. The confirmation of this phenomenon one can find in Fig. 2 where results on the decay of the electret potential are presented. The time constant of the decay increases with the applied voltage, due to the lower apparent conductivity of the polarized samples. The decrease in conductivity, as it is clear from Fig. l and 2 corresponds on the field of about 50-80 MV/m, the value being very close to the coercive field in PVDF [12]. It means that the abrupt decrease in conductivity is caused, most probably, by deep trapping of the charge carriers associated with the formation of the ferroelectric polarization.

## 3. Discussion and conclusions

The hypothesis on the deep trapping of the charge carriers in ferroelectric polymers is confirmed by our results, but the detailed mechanism of the charge trapping and its influence on stability of the polarization are not clearly understood.

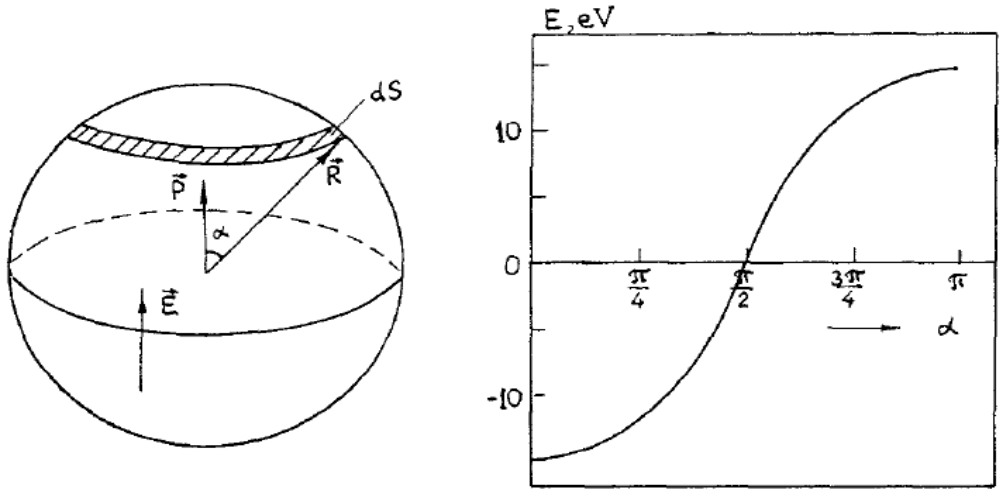

Fig. 3 Polarized spherical particle and distribution of the potential energy on its surface

We believe that the trapping may occur not only at permanent local centers, but also at potential wells created by the large-scale fluctuations of the potential energy. Let us consider a possibility of such processes in a ferroelectric polymer. If a spherical crystallite is uniformly polarized, the potential energy of the elementary charge on its surface is defined by the interaction between the charge and the local field induced by the polarized crystal

$$E = -\frac{1}{3\varepsilon_o} e \cdot P \cdot R \cdot \cos\alpha \qquad (1)$$

where $\varepsilon_o$ is the permittivity of free space, $e$ is the elementary charge, $P$ is the magnitude of the polarization vector, $R$ is the radius-vector, $\alpha$ is the angle between vectors $P$ and $R$.

It is clear from Eq. (1) and Fig. 3 that the potential energy is negative at a half of the surface, so conditions are favorable there for localization of the charge carriers. Suppose $n$ is the surface density of the localization sites and $g(E)$ is the energy distribution function of the traps. Then

$$g(E)dE = ndS \tag{2}$$

where $dS = 2\pi R^2 \sin\alpha d\alpha$

It follows from Eqs. (1) and (2) that

$$g(E) = \frac{3\varepsilon_o N}{ePR} = const \tag{3}$$

where $N$ is the total number of traps ($N = 2\pi R^2 n$).

In the range between 0 and $E_{min} = -\frac{ePR}{3\varepsilon_o}$, there is a uniform energy distribution of the localized states on the surface of the spherical polarized crystallites. For quantitative evaluation, we consider $V=10^{-19}$ cm$^3$ (R=30 Å). Substituting $P=0.14$ C/m$^2$ [12] we obtain $E$=15 eV, indicating that the carriers can be easily trapped at the surface of a crystallite.

In the case of crystallites in a form of plane discs of thickness $h$, which is very close to reality, one can get

$$E = -\frac{1}{4\varepsilon_o} e \cdot h \cdot P \cdot \cos\alpha \tag{4}$$

Substituting $h$=10Å and $P$=0.14 C/m$^2$ one obtains $E_{min}$=-4 eV that is also quite enough for deep trapping of the carriers.

Thus, during poling of the ferroelectric polymer, new traps are added to already existing Andersen's centers of localization in amorphous phase and to Maxwell-Wagner's traps at the crystallite boundaries. As a result, the thermodynamic equilibrium between delocalized and trapped carriers shifts to the direction of the latter. This affects the apparent mobility of carriers and, consequently, the apparent conductivity.

If the direction of the vector $P$ is reversed by an external field, the energy of the previously localized charge carriers will be drastically changed. As a result, the charge carriers are released from the traps, while other traps are formed on opposite sides of the crystallites. Although the total number of the traps and their energy distribution remain almost the same, their spatial distribution is different. Intensive release of carriers from the traps can be considered as a specific internal emission. During the definite time, there is a transient period of the abnormally high apparent conductivity due to the increased density of the free carriers.

In conclusion, the trapped charges play an important role in processes of stabilization and relaxation of the polarized state, since they compensate local depolarization fields. It is necessary to distinguish between the charge trapping in ferroelectric polymers and the charge accumulation due to Maxwell-Wagner's effect observed in all heterogeneous dielectrics, since the process in the former case is non-linear and irreversible. Moreover, the polarization in crystallites and trapped charges on their surfaces remain in a stable self-balanced state even after removal of the external field. We believe that similar processes take place in polymer-ceramics ferroelectric composites.